\documentclass[10pt,conference]{IEEEtran}
\usepackage{epsfig}
\usepackage{graphicx}
\usepackage{psfrag}

\begin{document}

\title{A Theory of Routing for  Large-Scale Wireless Ad-Hoc Networks}

\author{
\authorblockN{Antonio J. Caama\~{n}o, {\it Member, IEEE},
Juan J. Vinagre,
 Mark Wilby and
Javier Ramos, {\it Member, IEEE}
}
\authorblockA{Dept. Signal Theory and Communications,\
ETSIT-URJC,
Fuenlabrada, SPAIN}
\authorblockA{Email: $\{\mathsf{antonio.caamano,juanjose.vinagre,mark.wilby,javier.ramos}\}$@urjc.es}
}

\maketitle

\begin{abstract}

In this work we develop a new theory to analyse the process of routing in large-scale ad-hoc wireless networks. We use a path integral formulation to examine the properties of the paths generated by different routing strategies in these kinds of networks. Using this theoretical framework, we calculate the statistical distribution of the distances between any source to any destination in the network, hence we are able to deduce a length parameter that is unique for each routing strategy. This parameter, defined as the {\it effective radius}, effectively encodes the routing information required by a node.  Analysing the aforementioned statistical distribution for different routing strategies,  we obtain a threefold result for practical Large-Scale Wireless Ad-Hoc Networks: 1) We obtain the distribution of the lengths of all the paths in a network for any given routing strategy, 2) We are able to identify ``good" routing strategies depending on the evolution of its effective radius as the number of nodes, $N$, increases to infinity, 3) For any routing strategy with finite effective radius, we demonstrate that, in a large-scale network, is equivalent to a random routing strategy and that its transport capacity scales as $\Theta\left(\sqrt{N}\right)$ bit-meters per second, thus retrieving the scaling law that Gupta and Kumar (2000) obtained as the limit for single-route large-scale wireless networks. 
\end{abstract}

\section{Introduction}

The need for efficient routing in wireless ad-hoc networks, has given birth to a plethora of routing protocols \cite{Karl:2005ws}.  Due to the particular characteristics of those networks, the criteria followed to design the routing algorithms are somewhat different from the ones used in traditional, wired or wireless, networks with infrastructure \cite{Chen:1998dx}. The optimisation criteria range from the minimisation of the number of hops to reach destination, number of retransmissions, energy efficiency \cite{Zhao:2005ws} or topological considerations \cite{Melodia:2005of}. With the advent of ``large-scale" wireless ad-hoc networks, those routing strategies have to be revisited with new constraints in mind, such as the scalability \cite{Iwata:1999xb} and the capability for self-organization \cite{Sohrab:2000xy}.

Key to understand the differences of large-scale and small-scale wireless ad-hoc networks are the scaling laws discovered for single-route large-scale wireless networks \cite{Gupta:2000qn}. 

In this work a scaling law for the transport capacity of a network was given in the large-cale regime when any source and destination nodes were able to establish a route. This scaling law reflects that the individual throughput of each node decays as $\Theta(N^{-1/2})$ bits per second, thus vanishing with a increasing number of nodes. This law precludes a random routing strategy as it is not scalable. The authors suggest that local routing cells should be used to optimize the problem locally. This law has been verified experimentally in highly-detailed network simulations based on the IEEE 802.11 DCF \cite{Fang:2004jp}. 

In this work we develop a theoretical framework that is oriented to evaluate the efficiency of routing protocols in a dense wireless ad-hoc networks. We characterise the routing strategies by means of the capability of directing the route from source to destination. To this end we define an {\it energy of the route} where this directivity is implied. This energy ranges from zero value in a random routing strategy to infinity for a routing strategy capable of resolving the shortest path from any source node to any destination node. We will focus in the study of the distribution of the distances, relative to the length of the routing path, for any source-destination pair of nodes, a distribution defined as the {\it End-to-End distribution}. We investigate the characteristics of any routing algorithm based on a universal measure which we define as the {\it effective radius} of the routing protocol.  

The study of both the effective radius and the moments of the {\it End-to-End distribution} of the routing process reveal some interesting questions as to the pertinence of the use of any routing protocol with finite effective radius in Dense wireless ad-hoc networks. Our theoretical analysis shows that routing protocols that fall in the former category will behave as a purely random walk routing protocol in dense enough wireless ad-hoc networks. 
The validity of our analysis is shown as, in the {\it dense} regime of the network, we recover the famous transport capacity results derived by Gupta and Kumar \cite{Gupta:2000qn}.

\subsection{Organization and Summary of Results}

The network model and the description of the theoretical techniques used in this work are described in Section \ref{sec:prob}. In Section \ref{sec:e2ed}  we obtain analytically the End-to-End distribution for a routing strategy which is entirely described by a directivity function. The results for the specific cases of random walk routing and optimal routing are also described in Section \ref{sec:e2ed}. The case of Large-Scale Ad-Hoc Wireless Networks is treated in Section \ref{sec:rout} where the moments of the End-to-End distribution are analyzed for each of the three routing strategies defined in this work. A definition of the effective radius is also found in this section. We finally present the conclusions of this work, future directions and the limit of applicability of our results in Section \ref{sec:conc}.

The main results of this work are: the approximate formula of the distribution of the distances between nodes source an destination, the analytical distinction among different routing strategies in the case of Large-Scale Wireless Ad-Hoc Network and the retrieval of the transport capacity results for the case of a Large-Scale Wireless Ad-Hoc Network randomly distributed with random sources and destination of packets and single routing paths \cite{Gupta:2000qn}.

\subsection{Related Work}
The distribution of distances between source and destination nodes has been calculated before \cite{Miller:2001lo}\cite{Levedeb:2004tz}. 
Both cited approaches are dependent on a two-dimensional geometry which is justifiable up to some extent. In this work we opt for a three-dimensional formulation of the problem in order not to restrict the topology analyzed. But we are aware that the dimensionality of the routing problem in Wireless Ad-Hoc Networks is not a well defined problem.

The analysis of the routing problem dependent on a length scale that characterizes the awareness of the distributed routing protocol of its environment is not original and has been used before in the work by Melodia {\it et al.}\cite{Melodia:2005of}. The authors of this work introduce a phenomenological quantity called ``Knowledge Range" represents the physical extent of the routing strategy up to which is capable of finding the shortest path.

 The use of random walks as an effective (or unique) strategy for routing in Large-Scale Ad-Hoc Networks has been suggested in some works \cite{Servetto:2002ot} \cite{Rezaei:2004va}. In these works, the common drive to use this strategy is the logical conclusion that effective distributed routing in a large-scale network is unfeasible as it would require solving an NP-complete problem \cite{Chen:1998dx}.

\section{Problem Statement}\label{sec:prob}

\subsection{Network Model}
\begin{figure}[tbh!]
\begin{center}
\psfragscanon
\psfrag{a}{$a$}
\psfrag{0}{$0$}
\psfrag{1}{$1$}
\psfrag{2}{$2$}
\psfrag{N}{$N$}
\psfrag{xN-x0}{$\mathbf{R}$}
\psfrag{x0}{$\mathbf{x}_{0}$}
\psfrag{xN}{$\mathbf{x}_{N}$}
\psfrag{Dx1}{$\Delta\mathbf{x}_{1}$}
\psfrag{Dx2}{$\Delta\mathbf{x}_{2}$}
\psfrag{DxN-1}{$\Delta\mathbf{x}_{N-1}$}
\includegraphics[width=\columnwidth]{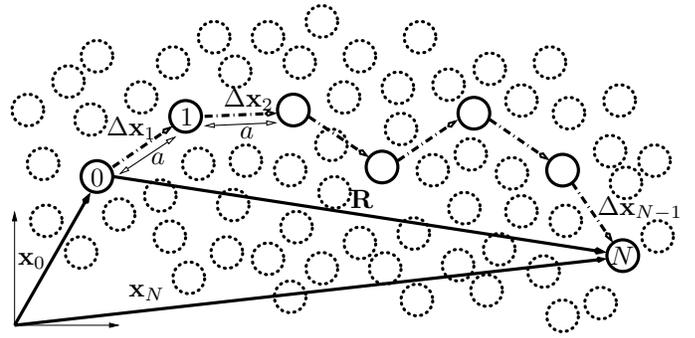}
\end{center}
\caption{Routing path in a wireless ad-hoc networks which is composed of $N$ links $\Delta\mathbf{x}_{n}$ of length $a$  connecting $\mathbf{x}_{0}$ and $\mathbf{x}_{N}$.}
\label{fig:net}
\end{figure}

The network model that we will use is summarized graphically in Figure \ref{fig:net}. The nodes of a wireless ad-hoc network are randomly distributed in a three-dimensional space. For the sake of clarity, graphical representations are shown in two dimensions even all the theoretical framework is formulated in three dimensions. The separation between the nodes  fluctuates around $a$ with a mean square $\left<\left(\Delta\mathbf{x}_{n}\right)^{2}\right>=a^{2}/3$. We formulate the routing process in terms of a chain of $N$ hops $\Delta\mathbf{x}_{n}$ between node $0$ (source) and node $N$ (destination).  The density of nodes is such that the radio coverage for each of the nodes allows the communications with neighboring nodes within the limitations of distances previously described. We will show later that, asymptotically, whether the distance between neighbouring nodes is fixed or fluctuating, the results are equal.

We assumed a perfect MAC in which every transmission is successful. This is done to establish a baseline comparison with respect to other works \cite{Gupta:2000qn}.
\subsection{Path Integral Method}

In order to obtain the aforementioned End-to-End distribution, we will use the {\it path integral} method \cite{Kleinert:2004sv}. In a very simplified way, this method was developed in the realm of quantum mechanics in order to obtain the probability amplitude that a quantum particle went from position $\mathbf{x}_{a}$ at time $\tau_{a}$ to position $\mathbf{x}_{b}$ at time $\tau_{b}$. The nature of quantum mechanics is such that, to  calculate correctly such probability amplitude, it is necessary to {\it sum aver all the possible histories}. We take advantage of this method to calculate the End-to-End distribution of all the possible routes with $N$ hops from node $0$ to node $N$

In quantum mechanics, the imaginary-time amplitude of a free particle with mass $M$ in natural units ($\hbar=1$) in three dimensions is 
\begin{equation}\label{eq:probamp}
(\mathbf{x}_{b}\tau_{b}|\mathbf{x}_{a}\tau_{a})=\frac{1}{\sqrt[3]{2\pi(\tau_b-\tau_a)/M}}\exp{\left[-\frac{M}{2}\frac{\mathbf{x}_b-\mathbf{x}_a}{\tau_b-\tau_a}\right]^2}
\end{equation}
 To obtain the End-to-End distribution of a Random Routing Strategy (RRS), we merely have to substitute $\mathbf{x}_b-\mathbf{x}_a$ by $\mathbf{R}$, $\tau_b-\tau_a$ by $Na$ and $M$ by $3/a$ and evaluate the path integral over all the possible routes
 \begin{equation}\label{eq:pathint}
P_{L}^{\mathsf{RRS}}(\mathbf{R})=\int\mathcal{D}^3 R\exp{\left\{-\frac{D}{2La}\int_{0}^{L}ds\left[\mathbf{R}(s)\right]^2\right\}}
\end{equation}
which in the case of route with $R^2<<Na^2$ equals to
\begin{equation}
P_{L}^{\mathsf{RRS}}(\mathbf{R})=\sqrt{\frac{3}{2\pi a}}e^{-3R^2/2La}
\end{equation}
where $L=Na$ is the length of the route. We will recover this result in Section \ref{sec:e2ed} through a more straightforward calculation, thus confirming its validity.

\section{End-to-end Distribution of Routing Strategies}\label{sec:e2ed}

\subsection{Directed Routing}
In the previous section we obtained a restricted analytical expression for the End-to-End distribution for a Random Routing Strategy. No specificity of the direction to go from one node to another was given. To be able to formulate a particular routing protocol we are to impose some form of directivity, some selection function in order to obtain a preference for a packet to hop to a given neighbour node is search for its destination,
We formulate the directionality of the hop from one node to the following in a simple expression that we will define as the {\it energy of the direction}. This energy will address the difference between the RRS (where the directionality was random) and the Directed Routing Strategy (DRS). 
To construct this functional, let us define a given route described by the one-parameter continuous function $\mathbf{r}(s)$. The tangent to that trajectory is given by by $\mathbf{u}(s)=\partial\mathbf{r}(s)/\partial s$. The variation of this tangent vector through the partial differential $\partial\mathbf{u}(s)/\partial s$. If we construct a quadratic functional from this variation of the ``direction" of the route, we will be able to calculate the energy stored by a particular route of length $L$ as 
\begin{equation}
E^{L}=\int_{0}^{L}\frac{\kappa}{2}\left(\frac{\partial \mathbf{u}(s)}{\partial s}\right)^{2}ds
\end{equation}
where $\kappa$ is defined as the {\it elastic constant}. 

The actual route is not continuous, $\partial\mathbf{u}(s)/\partial s$ being substituted by the difference of the vectors that link two consecutive pair of nodes $n$ and $n+1$. The actual form of the energy should be
\begin{equation}\label{eq:ener}
E^{N}_{\mathsf{DRS}}=\frac{\kappa}{2a}\sum_{n=1}^{N}\left(\mathbf{u}_{n}-\mathbf{u}_{n-1}\right)^{2}
\end{equation}

With that functional of the directivity defined, the end-to-end distribution of the DRS over a distance 
\begin{equation}
\mathbf{R}\equiv\mathbf{x}_{b}-\mathbf{x}_{a}=a\sum_{n=1}^{N}\mathbf{u}_{n}
\end{equation}
 is  obtained from the path integral with specific directions of the initial and final pieces 
\begin{eqnarray}\label{eq:e2eduaub}
P_{N}(\mathbf{u}_{b},\mathbf{u}_{a};\mathbf{R})&=&\frac{1}{A}\prod_{n=1}^{N-1}\left[\int\frac{d^{2}\mathbf{u}_{n}}{A}\right]\delta^{3}(\mathbf{R}-a\sum_{n=1}^{N}\mathbf{u}_{n})\nonumber\\
&\times&\exp\left[-\frac{2\pi a}{A^{2}}\sum_{n=1}^{N}\left(\mathbf{u}_{n}-\mathbf{u}_{n-1}\right)^{2}\right]
\end{eqnarray}
where $A$ is a measure factor given by
\begin{equation}
A=\sqrt{\frac{2\pi a}{\kappa\beta}}
\end{equation}

To obtain the desired end-to-end distribution, we integrate \ref{eq:e2eduaub} over all initial directions and average over the initial ones to obtain
is given by
\begin{equation}\label{eq:e2e-int} 
P_{N}^{\mathsf{DRS}}(\mathbf{R})=\int d^{2}\mathbf{u}_{b}\int \frac{d^{2}\mathbf{u}_{a}}{4\pi}P_{N}(\mathbf{u}_{b},\mathbf{u}_{a};\mathbf{R})
\end{equation}

The former equation is quite difficult to evaluate and, in general, the expressions of the End-to-End distributions are quite difficult to obtain analytically. Therefore, we shall work with the moments of the distribution instead of the distribution itself, which are found more easily. Thus, the moments of the distributions can be written as

\begin{equation}
\left<R^{2l}\right>=\int d^{2}\mathbf{u}_{b}\int \frac{d^{2}\mathbf{u}_{a}}{4\pi}R^{2l}P_{N}(\mathbf{u}_{b},\mathbf{u}_{a};\mathbf{R})
\end{equation}
We will take care only of the even moments of the distribution, as its dependence on the actual distance is  rotationally invariant. Therefor, odd moments vanish.  

If we introduce the angular distribution of a random chain of length $L=Na$ as

\begin{eqnarray}\label{eq:angul}
P_{N}(\mathbf{u}_{b},\mathbf{u}_{a}|L)&=&\frac{1}{A}\prod_{n=1}^{N-1}\left[\int\frac{d^{2}\mathbf{u}_{n}}{A}\right]\nonumber\\
&\times&\exp\left[-\frac{2\pi a}{A^{2}}\sum_{n=1}^{N}\left(\mathbf{u}_{n}-\mathbf{u}_{n-1}\right)^{2}\right]
\end{eqnarray}

We will be able to calculate the trivial moment $l=0$ (which will provide us with the proper normalization of the distribution) as
\begin{equation}
\left<1\right>=\int d^2 \mathbf{u}_b\int\frac{d^2\mathbf{u}_a}{4\pi}P_{N}(\mathbf{u}_{b},\mathbf{u}_{a}|L)=1
\end{equation}

Equation \ref{eq:angul} can be solved analytically 
\begin{equation}
P_{N}(\mathbf{u}_{b},\mathbf{u}_{a}|L)=\sum_{l=0}^{\infty}\exp{\left[-L\frac{1}{2\kappa\beta}L_{2}\right]}\sum_{\mathbf{m}}Y_{l\mathbf{m}}(\mathbf{u}_b)Y_{l\mathbf{m}}^{*}(\mathbf{u}_a)
\end{equation}
where $L_{2}=l(l+1)$ and $Y_{l\mathbf{m}}(\mathbf{u}_n)$ represent the harmonic polynomials.
Using  the orthogonality properties of the harmonic polynomials and rewriting the former integral in terms of Gegenbauer polynomials \cite{Gradshteyn:1994tb} and using their recursion relations, we are able to obtain the first nontrivial moment of the End-to-End distribution (details of the exact calculation will be published elsewhere).
Before writing the expression of $\left<R^{2l}\right>$ let us define the following quantity
\begin{equation}
\xi\equiv\kappa\beta
\end{equation}
which we call the {\it persistence radius}. The, the exact expression of the first nontrivial moment of the End-to-End distribution can be written as
\begin{equation}\label{eq:mom2}
\left<R^{2}_{\mathsf{DRS}}\right>=2\left[\xi L-\xi^2\left(1-e^{-L/\xi}\right)\right]
\end{equation}

Additional moments are increasingly difficult to calculate.

\subsection{Random Routing}
For Random Routing Strategy, we have previously calculated the End-to-End distribution with the path integral, but to validate such approach and to give further insight on the results, we will calculate it with a different approach. Let us have the route that we defined in Section \ref{sec:prob}, but instead of fluctuating, we will have a fixed distance between $a$ between consecutive nodes. If we have no preferred angle of direction to hop from one node to the following, we will have a Random Routing Strategy.
In three dimensions, the probability distribution of the end-to-end distance vector $\mathbf{x}_{b}-\mathbf{x}_{a}$ of such an route is given by
\begin{eqnarray}\label{eq:PNe2e}
P_{N}(\mathbf{R})&=&\prod_{n=1}^{N}\left[\int d^{3}\Delta x_{n}\frac{1}{4\pi a^{2}}\delta\left(|\Delta\mathbf{x}_{n}|-a\right)\right]\nonumber\\
                            				    &\times&\delta^{3}(\mathbf{R}-\sum_{n=1}^{N}\Delta\mathbf{x}_{n})
\end{eqnarray}
If we look at  equation \ref{eq:PNe2e} in terms of the Fourier transform of the one-link probabilities $\tilde{P}_{1}(\mathbf{k})$, we will obtain the desired integral as
\begin{eqnarray}
P_{N}(\mathbf{R})&=&\int\frac{d^{3}k}{(2\pi^{3})}\left[\tilde{P}_{1}(\mathbf{k})\right]^{N}e^{i\mathbf{kR}}\nonumber\\
			    &=&\frac{1}{2\pi^{2}R} \int_{0}^{\infty} dk k \sin kR\left[\frac{\sin ka}{ka}\right]^{N}
\end{eqnarray}
If we solve the previous integral, we should find that 
\begin{equation}
P_{L}^{\mathsf{RRS}}(\mathbf{R})=\sqrt{\frac{3}{2\pi a}}e^{-3R^2/2La}
\end{equation}
as found in the previous Section (we changed $N$ subscript by $L=Na$).

If we express the Fourier transform of $P_{N}(\mathbf{R})$ in terms of the moments of the end-to-end distribution of the RRP, we obtain
\begin{equation}
\tilde{P}_{N}(\mathbf{k})=\sum_{l=0}^{\infty}\frac{(-1)^{l}(k)^{2l}}{(2l)!}\frac{1}{2l+1}\left<R^{2l}\right>
\end{equation}
where the moments are 
\begin{equation}\label{eq:momrrs}
\left<R^{2l}_{\mathsf{RRS}}\right>=a^{2l}(-1)^l(2l+1)!\sum_{m_i}\prod_{i=1}^{l}\frac{1}{m_{i}!}\left[\frac{N2^{2i}(-1)^i B_{2i}}{(2i)!2i}\right]^{m_i}
\end{equation}
where the sum over $m_i$ obeys the constraint $l=\sum_{i=1}^{l}i\cdot m_i$ and $B_{i}$ are the Bernouilli numbers.

\subsection{Optimal Routing}
In a Optimal Routing Strategy (ORS), the packet is able to find the shortest path from source to destination, thus establishing a straight line between source and destination.
Its End-to-End distribution is trivial to find as 
\begin{equation}
P_{L}^{\mathsf{ORS}}(\mathbf{R})=\frac{1}{4\pi R^2}\delta(R-L)
\end{equation}
from which is straightforward to find that the moments of the ORS are
\begin{eqnarray}
\left<R^{n}_{\mathsf{ORS}}\right>&=&\int d^3 R R^n P_{L}^{\mathsf{ORS}}(\mathbf{R})\nonumber\\
&=&\int_{0}^{\infty}d^3 R R^n \delta(R-L)=L^n
\end{eqnarray}

\section{Routing in Large-Scale Ad-Hoc Wireless Networks}\label{sec:rout}
\subsection{Effective Radius of RS and the Large-Scale Limit }

Let us now revisit the moments of the End-to-End distribution for the three different routing strategies examined here (DRS, RRS and ORS) in the limit of a large number of nodes $N$ at finite $a^2 N$.

\begin{figure}[tbh!]
\begin{center}
\psfragscanon
\psfrag{a}{$a$}
\psfrag{0}{$0$}
\psfrag{1}{$1$}
\psfrag{2}{$2$}
\psfrag{N}{$N$}
\psfrag{p1}{$P$}
\psfrag{epsilon}{$a_{\mathsf{eff}}=1$}
\psfrag{epsilon1}{$a_{\mathsf{eff}}=2$}
\includegraphics[width=0.9\columnwidth]{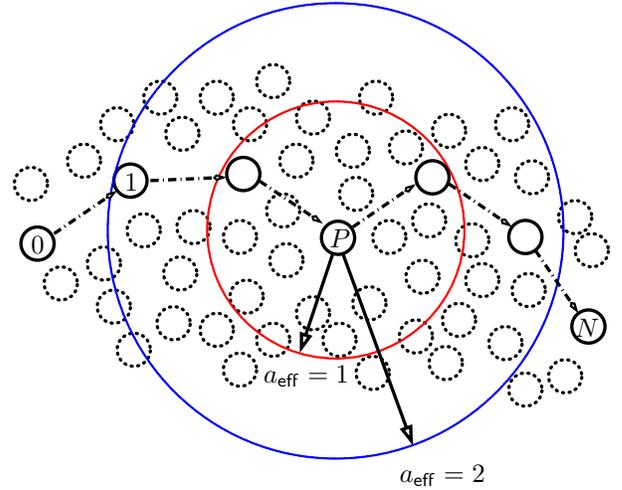}
\end{center}
\caption{Different effective radiuses $a_{\mathsf{eff}}$ of different routing protocols shown as radius of influence of a particular node.}
\label{fig:rad}
\end{figure}

We first take this limit in Equation \ref{eq:momrrs}, where we find
\begin{equation}\label{eq:momrrs2}
\left<R^{2l}_{\mathsf{RRS}}\right>=\frac{(2l+1)!!}{3^l}(aL)^{l}
\end{equation}
If we continue with Equation \ref{eq:mom2}, we can see that, in the limit for large $L/\xi$ we have
\begin{equation}
\left<R^{2}_{\mathsf{DRS}}\right>=2\xi L\left(1-\frac{\xi}{L}+\cdots\right)
\end{equation}
We can pursue a tedious calculation of the fourth moment of the DRS End-to-End distribution and, after the limits are taken, we have
\begin{equation}
\left<R^{4}_{\mathsf{DRS}}\right>=4\frac{5}{3}\xi^2 L^2\left(1-2\frac{26}{15}\frac{\xi}{L}+\dots\right)
\end{equation}
We can observe that the first term in both the second and the fourth moment are in accord with Equation \ref{eq:momrrs2} but, instead of a distance between nodes of $a$, we have 
an {\it effective radius} of the DRS 
\begin{equation}
a_{\mathsf{eff}}\equiv 2\xi
\end{equation} 
So, in the large-scale limit, we can see that the Directed Routing Strategy can be seen as a Random Routing Strategy (Figure \ref{fig:rad})  with a greater extent of influence.
As a consequence, we can rewrite the general expression for the moments of the DRP as, approximately
\begin{equation}
\left<R^{2l}_{\mathsf{DRS}}\right>\approx\frac{(2l+1)!!}{3^l}(a_{\mathsf{eff}}L)^l
\end{equation}

Finally we rewrite the Optimal Routing Strategy moments of the End-to-End Distribution as
\begin{equation}
\left<R^{2l}_{\mathsf{ORS}}\right>=L^{2l}
\end{equation}

\begin{figure}[tbh!]
\begin{center}
\psfragscanon
\psfrag{0}{$0$}
\psfrag{5}{$5$}
\psfrag{10}{$10$}
\psfrag{15}{$15$}
\psfrag{20}{$20$}
\psfrag{0.2}{$0.2$}
\psfrag{0.4}{$0.4$}
\psfrag{0.6}{$0.6$}
\psfrag{0.8}{$0.8$}
\psfrag{1}{$1$}
\psfrag{10-3}{$\xi/L=10^{-3}$}
\psfrag{10-1}{$\xi/L=10^{-1}$}
\psfrag{l=0.2}{$\xi/L=0.2$}
\psfrag{0.5}{$\xi/L=0.5$}
\psfrag{3}{$\xi/L=3$}
\psfrag{8}{$\xi/L=8$}
\psfrag{PR}[bl][l][1][90]{$P(R)$}
\psfrag{RL}{$R/L$}
\includegraphics[width=\columnwidth]{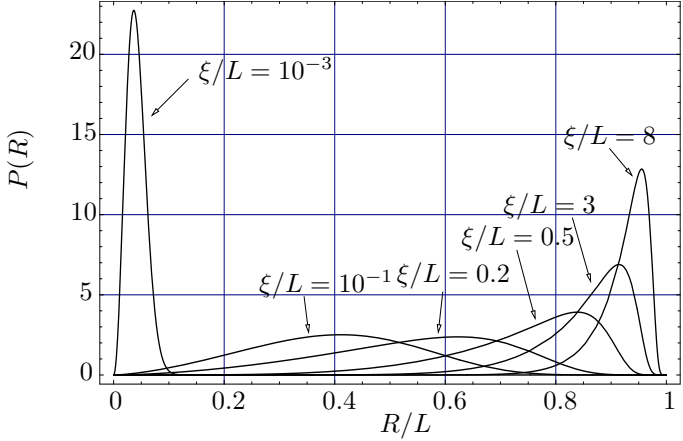}
\end{center}
\caption{End-to-End Distribution of a 3D Network for Routing Protocols with different persistence radiuses $\xi=a_{\mathsf{eff}}/2$.}
\label{fig:end-to-end distribution}
\end{figure}

So we can see a transition from a Random Routing Strategy whose influence extends only to its nearest neighbours, to a Random Routing Strategy but with a greater topological influence to, finally, an Optimal Routing Strategy whose influence extends to the whole of the network. 
To illustrate the distribution of the End-to-End distances with respect to the actual length of the routes, we have calculated numerically the End-to-End distribution for Routing Strategies with different effective radiuses (see Figure \ref{fig:end-to-end distribution}).

We can summarize the behaviour of the routing strategies as a whole by building the following moments function
\begin{equation}
\left<R^{2l}\right>\propto \left(a L\right)^{2l\nu}
\end{equation}
We will define $\nu$  as the {\it critical exponent}. This critical exponent would range from $\nu=1/2$ (Random Walk) to $\nu=1$ (Shortest Path).

In Figure \ref{fig:end-to-end distribution} we can clearly see that the distribution of distances between source and destination nodes for RRSs lie to the left of the figure ($\nu\approx 1/2$). The actual distance between the pair of nodes is negligible compared to the length of the route $L$. On the other hand, near-optimal DRSs (quasi-ORSs) lie to the right of the figure ($\nu\approx 1$). The distance between the pair of nodes is comparable and almost equal to the path of the route, thus assessing the optimality of the routing strategy. 

\subsection{Random Routing and Transport Capacity}

If we repeated the analysis done in this work for two dimensions, the result would be the same as to the functional of the moments of the End-to-End distribution, but with different multiplicative constants (the details of calculation will be published elsewhere). 
If we use a routing strategy with constant finite effective radius $a_{\mathsf{eff}}$ (e.g. a table-driven routing protocol), as we go to the large-scale limit is easy to see that 
\begin{equation}
\lim_{N\rightarrow\infty}\left<R^{2l}_{DRS}\right>=\left<R^{2l}_{RRS}\right>
\end{equation}
Therefore, as the network increases the number of nodes, any routing strategy which has less that total knowledge of the network available to each node, it will become a random routing strategy. In that limit, any packet originating from any node has a nonvanishing probability of visiting each node in the network. We are in an ergodic system. In two dimensions, it means that the average length of the routing path scales with the square root of the number of nodes, i.e.
\begin{equation}
\left<L\right>\propto N^{1/2}
\end{equation}
In three dimensions, however, the average length of the routing path scales as 
\begin{equation}
\left<L\right>\propto N^{3/2}
\end{equation}
Therefore, the transport capacity for a Large-Scale Ad-Hoc Wireless Network will obey a scaling law $\Theta\left(N^{1/2}\right)$ bits-meter per second in two dimensions and $\Theta\left(N^{3/2}\right)$ in three dimensions.

\section{Conclusions}\label{sec:conc}
In this work we introduced a novel method to analyze theoretically the routing strategies that are to be used in Large-Scale Wireless Ad-Hoc Networks. We have found the distribution of the distances between any source and destination nodes relative to the actual length of the routing path. We have shown that in the large-scale limit, any routing strategy will behave as a Random Walk. We have shown that the transport capacity of such large-scale networks are in agreement with those published in previous works. And, finally, we have deduced a new transport capacity depending on the dimensionality of the network.
\section*{Acknowledgments}
We would like to acknowledge useful discussions with  Javier Sim\'{o} Reigadas and Alfonso Cano Pleite.


\bibliographystyle{IEEEtran.bst}
%

\end{document}